\documentclass[11pt]{llncs}
\usepackage{t1enc}

\usepackage{pgfplots}
\usepackage{tikz}
\usepackage{fullpage}     
\usepackage{amsfonts}

\newcommand{\set}[1]{\left\{ #1\right\}}
\newcommand{\gilt}{:}
\newcommand{\sodass}{\,:\,}
\newcommand{\setGilt}[2]{\left\{ #1\sodass #2\right\}}

\newcommand{\realrange}[2]{\left[#1, #2\right]}

\newcommand{\unitrange}[2]{\realrange{0}{1}}

\newcommand{\llabel}[1]{\label{\labelprefix:#1}}
\newcommand{\labelprefix}{} 

\newcommand{\discussionsize}{\small}

\newcommand{\frage}[1]{}

\newenvironment{code}{\noindent
\begin{tabbing}%
\hspace{2em}\=\hspace{2em}\=\hspace{2em}\=\hspace{2em}\=\hspace{2em}\=%
\hspace{2em}\=\hspace{2em}\=\hspace{2em}\=\hspace{2em}\=\hspace{2em}\=%
\kill}{\end{tabbing}}

\newcommand{\labelcommand}{}
\newcommand{\captiontext}{}
\newsavebox{\codeparam}
\newcounter{lineNumber}
\newenvironment{disscodepos}[3]{%
\renewcommand{\labelcommand}{#2}%
\renewcommand{\captiontext}{#3}%
\sbox{\codeparam}{\parbox{\textwidth}{#3}}%
\begin{figure}[#1]\begin{center}\begin{code}\setcounter{lineNumber}{1}}{%
\end{code}\end{center}\caption{\llabel{\labelcommand}\captiontext}\end{figure}}

{\end{disscodepos}}

\newcommand{\Is}       {:=}

\newdimen\endofsize\endofsize=0.5em
\def\endofbeweis{~\quad\hglue\hsize minus\hsize
                 \hbox{\vrule height \endofsize width
\endofsize}\par}

\usepackage{epsfig}
\usepackage{url}
\usepackage{amsmath}
\usepackage{amssymb}
\usepackage{pdflscape}
\usepackage{latexsym}
\usepackage{todonotes}
\usepackage{hyperref}
\usepackage{algorithmic}
\usepackage{algorithm}
\usepackage{multicol}
\usepackage{wrapfig}
\usepackage{booktabs}
\usepackage{numprint}
\npdecimalsign{.} 
\usepackage{theorem}
\usepackage{makeidx}

\newcommand{\ie}{i.e.\ }
\newcommand{\etal}{et~al.\ }
\newcommand{\eg}{e.g.\ }

\newcommand{\Id}[1]{\texttt{\detokenize{#1}}}
\usepackage{color}
\usepackage{etoolbox}

\patchcmd{\thebibliography}{\list}{\fontsize{0.98em}{0.9\baselineskip}\selectfont\list}{}{} 
\newcommand{\csch}[1]{\color{orange}{\small [CS: #1]}}
\renewcommand{\csch}[1]{}

\newcommand{\mytitle}{Advanced Multilevel Node Separator Algorithms}
\begin{document}
\title{\mytitle}
\author{Peter Sanders and Christian Schulz\\ 
	\textit{Karlsruhe Institute of Technology},
	\textit{Karlsruhe, Germany} \\
\small	\textit{\{\url{sanders, christian.schulz}\}\url{@kit.edu}}}
\date{}
\institute{}

\maketitle
\begin{abstract}
A node separator of a graph is a subset $S$ of the nodes such that removing $S$ and its incident edges divides the graph into two disconnected components of about equal size.
In this work, we introduce novel algorithms to find small node separators in large graphs. 
With focus on solution quality, we introduce novel flow-based local search algorithms which are integrated in a multilevel framework. 
In addition, we transfer techniques successfully used in the graph partitioning field. 
This includes the usage of edge ratings tailored to our problem to guide the graph coarsening algorithm as well as  highly localized local search and iterated multilevel cycles to improve solution quality even further. 
Experiments indicate that flow-based local search algorithms on its own in a multilevel framework are already \emph{highly} competitive in terms of separator quality. 
Adding additional local search algorithms further improves solution quality.
Our strongest configuration almost always outperforms competing systems while on average computing 10\% and 62\% smaller separators than Metis and Scotch, respectively.

\end{abstract}
\thispagestyle{empty}
\section{Introduction}
Given a graph $G=(V,E)$, the \emph{node separator problem} asks to find three disjoint subsets $V_1, V_2$ and $S$ of the node set, such that there are no edges between $V_1$ and $V_2$ and $V=V_1\cup V_2 \cup S$. 
The objective is to minimize the size of the separator $S$ or depending on the application the weight of its nodes while $V_1$ and $V_2$ are balanced. 
Note that removing the set $S$ from the graph results in at least two connected components.
There are many algorithms that rely on small node separators. 
For example, small balanced separators are a popular tool in divide-and-conquer strategies~\cite{lipton1980applications,leiserson1980area,BHATT1984300}, are useful to speed up the computations of shortest paths~\cite{schulz2002using,delling2009high,dibbelt2014customizable} or are necessary in scientific computing to compute fill reducing orderings with nested dissection algorithms~\cite{george1973nested}.

Finding a balanced node separator on general graphs is NP-hard even if the maximum node degree is three~\cite{bui1992finding,garey2002computers}. 
Hence, one relies on heuristic and approximation algorithms to find small node separators in general graphs.
The most commonly used method to tackle the node separator problem on large graphs in practice is the multilevel approach.
During a coarsening phase, a multilevel algorithm reduces the graph size by iteratively contracting nodes and edges until the graph is small enough to compute a node separator by some other algorithm. A node separator of the input graph is then constructed by successively transferring the solution to the next finer graph and applying local search algorithms to improve the current solution. 

Current solvers are typically more than fast enough for most applications (for example~\cite{leiserson1980area,BHATT1984300}) but lack high solution quality.  In this work, we address this problem and focus on solution quality. The remainder of the paper is organized as follows.
We begin in Section~\ref{s:preliminaries} by introducing basic concepts and by summarizing related work.
Our main contributions are presented in Section~\ref{s:mainpartseparator} where we transfer techniques previously used for the graph partitioning problem to the node separator problem and introduce novel flow based local search algorithms for the problem that can be used in a multilevel framework. This includes edge ratings to guide a graph coarsening algorithm within a multilevel framework, highly localized local search to improve a node separator and iterated multilevel cycles to improve solution quality even further. 
Experiments in Section~\ref{s:experiments} indicate that our algorithms are able to provide excellent node separators and outperform other state-of-the-art algorithms.
Finally, we conclude with Section~\ref{s:conclusion}.
All of our algorithms have been implemented in the open source graph partitioning package KaHIP~\cite{kabapeE} and will be available within this framework.

\section{Preliminaries}
\label{s:preliminaries}
\subsection{Basic concepts}
In the following we consider an undirected graph $G=(V=\{0,\ldots, n-1\},E)$ with $n = |V|$, and $m = |E|$.
$\Gamma(v)\Is \setGilt{u}{\set{v,u}\in E}$ denotes the neighbors of a node $v$.
A set $C \subset V$ of a graph is called \textit{closed node set} if there are no connections from $C$ to $V \setminus C$, i.e. for every node $u \in C$ an edge $(u,v) \in E$ implies that $v \in C$ as well.
In other words, a subset $C$ is a \emph{closed node set} if there is no edge starting in $C$ and ending in its complement $V \setminus C$. 
A graph $S=(V', E')$ is said to be a \emph{subgraph} of $G=(V, E)$ if $V' \subseteq V$ and $E' \subseteq E \cap (V' \times V')$. We call $S$ an \emph{induced} subgraph when $E' = E \cap (V' \times V')$.  
For a set of nodes $U\subseteq V$, $G[U]$ denotes the subgraph induced by $U$.
We define multiple partitioning problems.
The \emph{graph partitioning problem} asks for \emph{blocks} of nodes $V_1$,\ldots,$V_k$ 
that partition $V$, i.e., $V_1\cup\cdots\cup V_k=V$ and $V_i\cap V_j=\emptyset$
for $i\neq j$. A \emph{balancing constraint} demands that 
$\forall i\in \{1..k\}\gilt |V_i|\leq L_{\max}\Is (1+\epsilon)\lceil |V|/k \rceil$ for
some parameter $\epsilon$. 
In this case, the objective is often to minimize the total \emph{cut} $\sum_{i<j}|E_{ij}|$ where 
$E_{ij}\Is\setGilt{\set{u,v}\in E}{u\in V_i,v\in V_j}$. 
The set of cut edges is also called \emph{edge separator}.
A node $v \in V_i$ that has a neighbor $w \in V_j, i\neq j$, is a boundary node. 
An abstract view of the partitioned graph is the so called \emph{quotient graph}, where nodes represent blocks and edges are induced by connectivity between blocks. 
The \emph{node separator problem} asks to find blocks, $V_1, V_2$ and a separator $S$ that partition $V$ such that there are no edges between the blocks. 
Again, a balancing constraint demands $|V_i| \leq (1+\epsilon)\lceil|V|/k \rceil $. However, there is no balancing constraint on the separator $S$. 
The objective is to minimize the size of the separator $|S|$. 
Note that removing the set $S$ from the graph results in at least two connected components and that the blocks $V_i$ itself do not need to be connected components.
By default, our initial inputs will have unit edge and node weights. However, the results in this paper are easily transferable to node and edge weighted problems.
A matching $M\subseteq E$ is a set of edges that do not share any common nodes,
\ie the graph $(V,M)$ has maximum degree one.  \emph{Contracting} an edge $\set{u,v}$ means to replace the nodes $u$ and $v$ by a
 new node $x$ connected
to the former neighbors of $u$ and $v$. We set $c(x)=c(u)+c(v)$.
If replacing
edges of the form $\set{u,w},\set{v,w}$ would generate two parallel edges
$\set{x,w}$, we insert a single edge with
$\omega(\set{x,w})=\omega(\set{u,w})+\omega(\set{v,w})$.
\emph{Uncontracting} an edge $e$ undos its contraction. 
In order to avoid tedious notation, $G$ will denote the current state of the graph
before and after a (un)contraction unless we explicitly want to refer to 
different states.

The multilevel approach consists of three main phases.
In the \emph{contraction} (coarsening) phase, 
we iteratively identify matchings $M\subseteq E$ 
and contract the edges in $M$. Contraction should quickly reduce the size of the input and each computed level
should reflect the global structure of the input network. 
Contraction is stopped when the graph is small enough so that the problem can be solved by some other potentially more expensive algorithm. 
In the \emph{local search} (or uncoarsening) phase, matchings are iteratively uncontracted.  
After uncontracting a matching, the local search algorithm moves nodes to decrease the size of the separator or to to improve balance of the block while keeping the size of the separator. 
The succession of movements is based on priorities called \textit{gain}, i.e., the decrease in the size of the separator.  
The intuition behind the approach is that a good solution at one level of the hierarchy will also be a good solution on the next finer level so that local search will quickly find a good solution.

\subsection{Related Work}
\label{s:related}
There has been a \emph{huge} amount of research on graph partitioning so that we refer the reader to \cite{GPOverviewBook,SPPGPOverviewPaper} for most of the material in this area. 
Here, we focus on issues closely related to our main contributions and previous work on the node separator problem. 
Lipton and Tarjan~\cite{lipton1979separator} provide the \textit{planar separator theorem} stating that on planar graphs one can always find a separator $S$ in linear time that satisfies $|S| \in O(\sqrt{|V|})$ and $|V_i| \leq 2 |V| / 3$. 
For more balanced cases, the problem remains NP-hard~\cite{fukuyama2006np} even on planar graphs. 

For general graphs there exist several heuristics to compute small node separators. 
A common and simple method is to derive a node separator from an edge separator \cite{pothen1990partitioning,dissSchulz} which is usually computed by a multilevel graph partitioning algorithm.
Clearly, taking the boundary nodes of the edge separator in one block of the partition yields a node separator. 
Since one is interested in a small separator, one can use the smaller set of boundary nodes.
A better method has been first described by Pothen and Fan~\cite{pothen1990partitioning}.
The method employs the set of cut edges of the partition and computes the smallest node separator that can be found by using a subset of the boundary nodes. 
The main idea is to compute a subset $S$ of the boundary nodes such that each cut edge is incident to at least one of the nodes in $S$ (a vertex cover). 
A problem of the method is that the graph partitioning problem with edge cut as objective has a somewhat different combinatorial structure compared to the node separator problem. This makes it unlikely to find high quality solutions with that approach. 

Metis~\cite{karypis1998fast} and Scotch~\cite{scotch} use a multilevel approach to obtain a node separator. 
After contraction, both tools compute a node separator on the coarsest graph using a greedy algorithm.
This separator is then transferred level-by-level, dropping non-needed nodes on each level and applying Fiduccia-Mattheyses (FM) style local search. 
Previous versions of Metis and Scotch also included the capability to compute a node separator from an edge separator.

Recently, Hamann and Strasser~\cite{hamann2015graph} presented a max-flow based algorithm specialized for road networks. 
Their main focus is not on node separators. They focus on a different formulation of the edge-cut version graph partitioning problem. More precisely, Hamann and Strasser find Pareto solutions in terms of edge cut versus balance instead of specifying the allowed amount of imbalance in advance and finding the best solution satisfying the constraint.
Their work also includes an algorithm to derive node separators, again in a different formulation of the problem, \ie node separator size versus balance. 
We cannot make meaningful comparisions since the paper contains no data on separator quality and the implementation of the algorithm is not available.

Hager \etal\cite{hager2014multilevel} recently proposed a multilevel approach for medium sized graphs using continuous bilinear quadratic programs and a combination of those with local search algorithms. However, a different formulation of the problem is investigated, \ie the solver enforces upper \emph{and} lower bounds to the block sizes which makes the results incomparable to our results. 

LaSalle and Karypis~\cite{lasalle2015efficient} present a shared-memory parallel algorithm to compute node separators used to compute fill reducing orderings.
Within a multilevel approach they evaluate different local search algorithms indicating that a combination of greedy local search with a segmented FM algorithm can outperform serial FM algorithms. We compare the solution quality of our algorithm against the data presented there in our experimental section (see Section~\ref{s:experiments}).

\vfill
\pagebreak
\section{Advanced Multilevel Algorithms for Node Separators}
\label{s:mainpartseparator}
We now present our core innovations. 
In brevity, the novelties of our algorithm include edge ratings during coarsening to compute graph hierarchies that fulfill the needs of the node separator problem and a combination of localized local search with flow problems to improve the size of the separator. In addition, we transfer a concept called iterative multilevel scheme previously used in graph partitioning to further improve solution quality. 
The description of our algorithm in this section follows the multilevel scheme. 
We start with the description of the edge ratings that we use during coarsening, continue with the description of the algorithm used to compute an initial node separator on the coarsest level and then describe local search algorithms as well as other techniques.
\subsection{Coarsening}
Before we explain the matching algorithm that we use in our system, we present the general two-phase procedure which was already used in multiple graph partitioning frameworks \cite{kappa,kaffpa,kaspar}.
The two-phase approach makes contraction more systematic by separating two issues: A \emph{rating function} and a \emph{matching} algorithm. 
A rating function indicates how much sense it makes to contract an edge based on \emph{local} information.  
A matching algorithm tries to maximize the sum of the ratings of the contracted edges looking at the \emph{global} structure of the graph. 
While the rating function allows a flexible characterization of what a ``good'' contracted graph is, the simple, standard definition of the matching problem allows to reuse previously developed algorithms for weighted matching. 
Note that we can use the same edge rating functions as in the graph partitioning case but also can define new ones since the problem structure of the node separator problem is different. 

We use the \textit{Global Path Algorithm (GPA)} which runs in near linear time to compute matchings. 
GPA was proposed in \cite{MauSan07} as a synthesis of the Greedy Algorithm and the Path Growing Algorithm~\cite{DH03a}. 
We choose this algorithm since in \cite{kappa} it gives empirically considerably better results than Sorted Heavy Edge Matching, Heavy Edge Matching or Random Matching \cite{SchKarKum00}.
GPA scans the edges in order of decreasing weight
but rather than immediately building a matching, it first constructs a collection
of paths and even length cycles. Afterwards, \textit{optimal solutions} are computed for each
of these paths and cycles using dynamic programming. 

\paragraph{Edge Ratings for Node Separator Problems.}
We want to guide the contraction algorithm so that coarse levels in the graph hierarchy still contain small node separators if present in the input problem. This way we can provide a good starting point for the initial node separator routine.
There are a lot of possibilities that we have tried. 
The most important edge rating functions for an edge $e=\{u,v\} \in E$ are the following:
\begin{align*}
       \text{exp*}(e)&=  \omega(e)/(d(u)d(v))  \\
       \text{exp{**}}(e) &=  \omega(e)^2/(d(u)d(v)) \\
       \text{max}(e) &=  1/\max\{{d(u),d(v)}\} \\
       \text{log}(e) &=  1/\log(d(u)d(v))
\end{align*}
The first two ratings have already been successfully used in the graph partitioning field. 
To give an intuition behind these ratings, we have to characterize the properties of ``good'' matchings for the purpose of contraction in a multilevel algorithm for the node separator problem. 
Our main objective is to find a small node separator on the coarsest graph. 
A matching should contain a \emph{large number of edges}, \eg being maximal, so that there are only few levels in the hierarchy and the algorithm can converge quickly. 
In order to \emph{represent the input} on the coarser levels, we want to find matchings such that the graph after contraction has somewhat \emph{uniform node weights} and \emph{small node degrees}. In addition, we want to keep nodes having a small degree since they are potentially good separators. 
Uniform node weights are also helpful to achieve a balanced node separator on coarser levels and makes local search algorithms more effective.
We also included ratings that do not contain the edge weight of the graph since intuitively a matching does not have to care about large edge weights -- they do not show up in the objective of the node separator problem. 
\subsection{Initial Node Separators}
We stop coarsening as soon as the graph has less than ten thousand nodes. 
Our approach first computes an edge separator and then derives a node separator from that. More precisely, we partition the coarsest graph into two blocks using KaFFPa~\cite{kabapeE}.
We then look at the bipartite graph induced by set of cut edges including the given node weights.
Our goal is to select a minimum weight node separator in that graph.
As a side note, this corresponds to finding a minimum weight vertex cover in the bipartite graph. 
Also note that this is similar to the approach of Pothen \etal~\cite{pothen1990partitioning}, however we integrate node weights.
To solve the problem, we put all of the nodes of the bipartite graph into the initial separator $S$ and use the \emph{flow-based technique} defined below to select the smallest separator contained in that subgraph. 
Since our algorithms are randomized, we repeat the overall procedure twenty five times and pick the best node separator that we have found. 
\subsection{Local Search}
\paragraph{Localized Local Search.}
In graph partitioning it has been shown that higher localization of local search can improve solution quality~\cite{dissSchulz,kaspar}. 
Hence, we develop a novel localized algorithm for the node separator problem that starts local search only from a couple of selected separator nodes. 
Our localized local search procedure is based on the FM scheme. 
Before we explain our approach to localization, we present a commonly used FM-variant for completeness. 

For each of the two blocks $V_1$, $V_2$ under consideration, a priority queue of separator nodes eligible to move is kept. 
The priority is based on the \emph{gain} concept, \ie the decrease in the objective function value when the separator node is moved into that block. 
More precisely, if a node $v \in S$ would be moved to $V_1$, then the neighbors of $v$ that are in $V_2$ have to be moved into the separator. 
Hence, in this case the gain of the node is the weight of $v$ minus the weight of the nodes that have to be added to the separator. 
The gain value in the other case (moving $v$ into to $V_2$) is similar. 
After the algorithm computed both gain values it chooses the largest gain value such that moving the node does not violate the balance constraint and performs the movement.
Each node is moved at most once out of the separator within a single local search. 
The queues are initialized randomly with the separator nodes. 
After a node is moved, newly added separator nodes become eligible for movement (and hence are added to the priority queues). 

There are different possibilities to select a block to which a node shall be moved. 
The most common variant of the classical FM-algorithm alternates between both blocks. 
After a stopping criterion is applied, the best feasible node separator found is reconstructed (among ties choose the node separator that has better balance). 
We have two strategies to \emph{balance blocks}. 
The first strategy tries to create a balanced situation without increasing the size of the separator. It always selects the queue of the heavier block and uses the same roll back mechanism as before. 
The second strategy allows to increase the size of the node separator. 
It also selects a node from the queue of the heavier block, but the roll back mechanism recreates the node separator having the best balance (among ties we choose the smaller node separator). 

Our approach to localization works as follows. 
Previous local search methods were initialized with \emph{all} separator nodes, \ie all separator nodes are eligible for movement at the beginning.
In contrast, our method is repeatedly initialized only with a \emph{subset} of the separator nodes (the precise amount of nodes in the subset is a tuning parameter).  
Intuitively, this introduces a larger amount of diversification and boosts the algorithms ability to climb out of local minima.

The algorithm is organized in rounds. 
One round works as follows. 
Instead of putting \emph{all} separator nodes directly into the priority queues, we put the current separator nodes into a todo list $T$. 
Subsequently, we begin local search starting with a random \emph{subset} $\mathcal{S}$ of the todo list $T$. 
We select the subset $\mathcal{S}$ by repeatedly picking a random node $v$ from $T$. 
We add $v$ to $\mathcal{S}$ if it still is a separator node and has not been moved by a previous local search in that round.
Either way, $v$ is removed from the todo list. 
Our localized search is restricted to the movement of nodes that have not been touched by a previous local search during the round.
This assures that each node is moved at most once out of the separator during a round of the algorithm and avoids cyclic local search. By default our local search routine first uses classic local search (including balancing) to get close to a good solution and afterwards uses localization to improve the result further.
We repeat this until no further improvement is found.

We now give intuition why localization of local search boosts the algorithms ability to climb out of local minima.
Consider a situation in which a node separator is a locally optimal in the sense that at least two node movements are necessary until moving a node out of the separator with positive gain is possible. Recall that classical local search is initialized with all separator nodes (in this case all of them have negative gain values). 
It then starts to move nodes with negative gain at multiple places of the graph. 
When it finally moves nodes with positive gain the separator is already much worse than the input node separator.
Hence, the movement of these positive gain nodes does not yield an improvement with respect to the given input partition. 
On the other hand, a localized local search that starts close to the nodes with positive gain, can find the positive gain nodes by moving only a small number of nodes with negative gain. 
Since it did not move as many negative gain nodes as the classical local search, it may still finds an improvement with respect to the input.
\paragraph{Maximum Flows as Local Search.}
We define the node-capacitated flow problem $\mathcal{F}=(V_\mathcal{F}, E_\mathcal{F})$ that we solve to improve a given node separator as follows.
First we introduce a few notations. 
Given a set of nodes $A \subset V$, we define its \emph{border}  
$\partial A := \{ u \in A \mid \exists (u,v) \in E : v \not\in A\}$.
The set $\partial_1 A := \partial A \cap V_1$ is called \emph{left border} of $A$ and the set $\partial_2 A := \partial A \cap V_2$ is called \emph{right border} of $A$. 
An \emph{$A$ induced flow problem} $\mathcal{F}$ is the node induced subgraph $G[A]$ using $\infty$ as edge-capacities and the node weights of the graph as node-capacities. Additionally there are two nodes $s,t$ that are connected to the border of $A$. 
More precisely, $s$ is connected to all left border nodes $\partial_1 A$ and all right border nodes $\partial_2 A$ are connected to $t$.  
These new edges get capacity $\infty$. 
Note that the additional edges are directed.
$\mathcal{F}$ has the \emph{balance property} if each ($s$,$t$)-flow induces a balanced node separator in $G$, \ie the blocks $V_i$ fulfill the balancing constraint.
The basic idea is to construct a flow problem $\mathcal{F}$ having the balance property. 
We now explain how we find such a subgraph.
We start by setting $A$ to $S$ and extend it by performing two breadth first searches (BFS). 
The first BFS is initialized with the current separator nodes $S$ and only looks at nodes in block $V_1$.
The same is done during the second BFS with the difference that we now look at nodes from block $V_2$.
Each node touched by any of the BFS is added to $A$.
The first BFS is stopped as soon as the size of the newly added nodes would exceed $L_\text{max}-c(V_2)-c(S)$. Similarly, the second BFS is stopped as soon as the size of the newly added nodes would exceed $L_\text{max}-c(V_1)-c(S)$. 
\begin{figure}[t]
\begin{center}
\begin{tabular}{c}
\includegraphics[width=230pt]{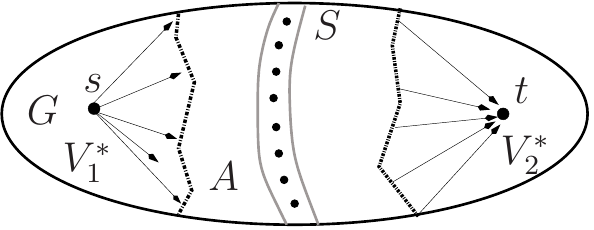}
\end{tabular}
\end{center}
\caption{ The construction of an $A$ induced flow problem $\mathcal{F}$ is shown. 
        Two breadth first searches are started to define the area $A$ -- one into the block on the left hand side and one into the block on the right hand side.
        A solution of the flow problem yields the smallest node separator that can be found within the area.  The area $A$ is chosen so that each node separator that can be found in the area yields a feasible separator for the original problem.}
        \label{fig:flowconstruction}
\end{figure}

A solution of the $A$ induced flow problem yields a valid node separator of the original graph:
First, since all edges in our flow network have capacity $\infty$ and the separator $S$ is contained in the problem, a maximum flow yields a separator $S'$, $V_\mathcal{F}=V'_1 \cup V'_2 \cup S'$, in the flow network that separates $s \in V'_1$ from $t \in V'_2$.
Since there is a one-to-one mapping between the nodes of our flow problem and the nodes of the input graph, we directly obtain a separator in the original network $V=V^*_1\cup V^*_2 \cup S'$.
Additionally, the node separator computed by our method fulfills the balance constraint -- presuming that the input solution is balanced.
To see this, we consider the size of $V^*_1$. We can bound the size of this block by assuming that all of the nodes that have been touched by the second BFS get assigned to $V^*_1$ (including the old separator $S$). 
However, in this case the balance constraint is still fulfilled $c(V^*_1) \leq c(V_1) + c(S) + L_\text{max} - c(V_1) - c(S) = L_\text{max}$.
The same holds for the opposite direction.
Note that the separator is always smaller or equal to the input separator since $S$ is contained in the construction.

To solve the node-capacitated flow problem $\mathcal{F}$, we transform it into a flow problem $\mathcal{H}$ without node-capacities.
We use a standard technique~\cite{ravindra1993network}: first we insert the source and the sink into our model. Then, for each node $u$ in our flow problem $\mathcal{F}$ that is not the source or the sink, we introduce two nodes $u_1$ and $u_2$ in $V_\mathcal{H}$ which are connected by a directed edge $(u_1,u_2) \in E_\mathcal{H}$ with an edge-capacity set to the node-capacity of the current node. 
For an edge $(u,v) \in E_\mathcal{F}$ not involving the source or the sink, we insert $(u_2, v_1)$ into $E_\mathcal{H}$ with capacity $\infty$.
If $u$ is the source $s$, we insert $(s,v_1)$ and if $v$ is the sink, we insert $(u_2, t)$ into $E_\mathcal{H}$. In both cases we use capacity $\infty$.

\paragraph{Larger Flow Problems and Better Balanced Node Separators.}
The definition of the flow problem to improve a node separator requires that each cut in the flow problem corresponds to a \emph{balanced} node separator in the original graph. We now simplify this definition and stop the BFSs if the size of the touched nodes exceeds $(1+\alpha) L_\text{max}-c(V_i)-c(S)$ with $\alpha \geq 0$. 
We then solve the flow problem and check afterwards if the corresponding node separator is balanced. If this is the case, we accept the node separator and continue. 
If this is not the case, we set $\alpha := \alpha/2$ and repeat the procedure. After ten unsuccessful iterations, we set $\alpha=0$. 
Additionally, we stop the process if the flow value of the flow problem corresponds to the separator weight of the input separator.
\begin{figure}[t]
\begin{center}
\includegraphics[width=0.6\textwidth]{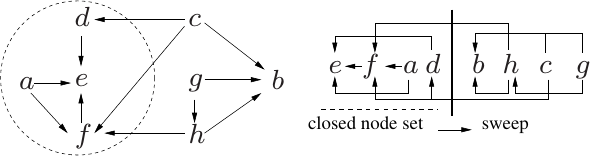}
\end{center}
\caption{Left: the set $C=\{a,d,e,f\}$ is a closed node set since no edge is starting in $C$ and ending in $V\backslash C$. Right: using a reverse topological ordering of a DAG one can output multiple closed node sets.}
\label{fig:closednodeset}
\end{figure}

We apply heuristics to extract a better balanced node separator from the solved max-flow problem.
Picard and Queyranne~\cite{picard1980structure} made the observation that \emph{one} $(s,t)$-max-flow contains information about \emph{all} minimum ($s$,$t$)-cuts in the graph (however, finding the most balanced minimum cut is NP-hard~\cite{bonsma2010most}).
We follow the heuristic approach of \cite{kaffpa} and extract better balanced ($s$,$t$)-cuts from the given maximum flow in $\mathcal{H}$. This results in better balanced separators in the node-capacitated problem $\mathcal{F}$ and hence in better balanced node separators for our original problem.

To be more precise, Picard and Queyranne have shown that each closed node set in the residual graph of a maximum $(s,t)$-flow that contains the source $s$ but not the sink  induces a minimum $s$-$t$ cut.
Observe that a cycle in the residual graph cannot contain a node of both, a closed node set and its complement.
Hence, Picard and Queyranne compactify the residual network by contracting all strongly connected components.
Afterwards, their algorithm tries to find the most balanced minimum cut by enumeration. 
In \cite{kaffpa}, we find better balanced cuts heuristically. 
First a random topological order of the strongly connected component graph is computed.
This is then scanned in reverse order. 
By subsequently adding strongly connected components several closed node sets are obtained, each inducing a minimum $s$-$t$ cut. 
The closed node set with the best occurred balance among multiple runs of the algorithm with different random topological orders is returned. 
An example closed node set and the scanning algorithm is shown in Figure~\ref{fig:closednodeset}.

\subsection{Miscellanea}
An easy way to obtain high quality node separators is to use a multilevel algorithm multiple times using different random seeds and use the best node separator that has been found.
However, instead of performing a full restart, one can use the information that has already been obtained.
In the graph partitioning context, the notion of iterated multilevel schemes has been introduced by Walshaw \cite{walshaw2004multilevel} and later has been augmented to more complex cycles~\cite{kaffpa}.  
Here, one transfers a solution of a previous multilevel cycle down the hierarchy and uses it as initial solution. 
More precisely, this can be done by not contracting any cut edge. 

We \emph{transfer this technique} to the node separator problem as follows. 
One can interpret a node separator as a three way partition $V_1,V_2,S$. 
Hence, to obtain an iterated multilevel scheme for the node separator problem, our matching algorithm is not allowed to match any edge that runs between $V_i$ and $S$ ($i=1,2$). 
Hence, when contraction is done, every edge leaving the separator will remain and we can transfer the node separator down in the hierarchy.
Thus a given node separator can be used as initial node separator of the coarsest graph (having the same balance and size as the node separator of the finest graph).  
This ensures non-decreasing quality, if the local search algorithm guarantees no worsening. 
To increase diversification during coarsening in later V-cycles we pick a random edge rating of the ones described above.

\vfill
\pagebreak
\section{Experiments}
\label{s:experiments}
\paragraph*{Methodology.} 
We have implemented the algorithm described above within the KaHIP framework using C++ and compiled all algorithms using gcc 4.63 with full optimization's turned on (-O3 flag). 
We integrated our algorithms in KaHIP v0.71 and compare ourselves against Metis~5.1 and Scotch~6.0.4 using the quality option that has focus on solution quality instead of running time. 
Our new codes will be included into the KaHIP graph partitioning framework.
We perform ten repetitions of each algorithm using different random seeds for initialization.
Each run was made on a machine that has four Octa-Core Intel Xeon E5-4640 processors running at 2.4\,GHz. 
It has 512 GB local memory, 20 MB L3-Cache and 8x256 KB L2-Cache.
Our main objective is the cardinality of node separators on the input graph. 
In our experiments, we use $\epsilon=20\%$ since this is the default value for node separators in Metis. 
We mostly present two kinds of views on the data: average values and minimum values as well as plots that show the ratios of the quality achieved by the algorithms. 

\paragraph*{Algorithm Configuration.}
We performed a number of experiments to evaluate the influence and choose the parameters of our algorithms.
We mark the instances that have also been used for the parameter tuning in Appendix~\ref{sec:appendixdetailedresults} with a * and exclude these graphs when we report average values over multiple instances in comparisons with our competitors. However, our full algorithm is not too sensitive about the precise choice with most of the parameters.
In general, using more sophisticated edge ratings improves solution quality slightly and improves partitioning speed over using edge weight. 
We exclude further experiments from the main text and use the $exp^*$ edge rating function as a default since it has a slight advantage in our preliminary experiments. 
In later iterated multilevel cycles, we pick one of the other ratings at random to introduce more diversification. 
Indeed, increasing the number of V-cycles reduces the objective function. 
We fixed the number of V-cycles to three. 
By default, we use the better balanced minimum cut heuristic in our node separator algorithm since it keeps the node separator cardinality and improves balance. In the localized local search algorithm, we set the size of the random subset of separator nodes from which local search is started $|\mathcal{S}|$ to five. 
\paragraph*{Instances.}
We use graphs from various sources to test our algorithm. 
We use all 34 graphs from Chris Walshaw's benchmark archive~\cite{soper2004combined}.
Graphs derived from sparse matrices have been taken from the Florida Sparse Matrix Collection~\cite{UFsparsematrixcollection}. 
We also use graphs from the 10th DIMACS Implementation Challenge~\cite{benchmarksfornetworksanalysis} website. 
Here, \Id{rggX} is a \emph{random geometric graph} with
$2^{X}$ nodes where nodes represent random points in the unit square and edges
connect nodes whose Euclidean distance is below $0.55 \sqrt{ \ln n / n }$.
The graph \Id{delX} is a Delaunay triangulation of $2^{X}$ random points in the unit square. 
The graphs \Id{af_shell9}, \Id{thermal2},  \Id{nlr} and \Id{nlpkkt240} are from the matrix and the numeric section of the DIMACS benchmark set.
The graphs \Id{europe} and \Id{deu} are large road networks of Europe and Germany taken from~\cite{DSSW09}. 
Due to large running time of our algorithm, we exclude the graph \Id{nlpkkt240} from general comparisons and only use our full algorithm to compute a result. 
Basic properties of the graphs under consideration can be found in Appendix~\ref{apdx:graphs}, Table~\ref{tab:test_instances_walshaw}.

\vfill
\pagebreak

\subsection{Separator Quality}
\setlength{\tabcolsep}{1ex}
\begin{wraptable}{r}{7cm}
\centering
\vspace*{-.75cm}
\begin{tabular}{lrrr}
\toprule
Algorithm & Avg. Inc.  & $t_\text{avg}$[s] & $\#\leq_\text{Metis}$ \\
\midrule
Metis          & 10.3\%  &0.12&-\\ 
Scotch         & 62.2\%  & 0.23& 0\%\\ 
\midrule
Flow$_0$       & 3.3\%  &17.72 & 89\%\\
Flow$_{0.5}$   & 0.1\%  &38.21 & 96\%\\
Flow$_1$       & 0.3\%  &47.81& 94\%\\
\midrule
LSFlow$_0$     & 1.5\%  &28.61&96\%\\
LSFlow$_{0.5}$ & -0.1\%  &49.08&94\%\\
\hline
\hline
LSFlow$_{1}$   & -  &58.50&96\%\\
                       \bottomrule
\end{tabular}
\caption{Avg. increase in separator size over LSFlow$_1$ , avg. running times of the different algorithms and relative number of instances with a separator smaller or equal to Metis ($\#\leq_\text{Metis}$).}
\vspace*{-.75cm}
\label{tab:compressedresults}
\end{wraptable}

We now assess the size of node separators derived by our algorithms and by other state-of-the-art tools, \ie Metis and Scotch as well as the data recently presented by LaSalle and Karypis~\cite{lasalle2015efficient}.
We use multiple configurations of our algorithm to estimate the influence of the multiplicative factor $\alpha$ that controls the size of the flow problems solved during uncoarsening and to see the effect of adding local search.
The algorithms named Flow$_\alpha$ use \emph{only} flows during uncoarsening as local search with a multiplicative factor $\alpha$. Algorithms labeled LSFlow$_\alpha$ start on each level with local search and localized local search until no improvement is found and afterwards perform flow based local search with a multiplicative factor $\alpha$.
Table~\ref{tab:compressedresults} summarizes the results of the experiments. We present detailed per instances results in Appendix~\ref{sec:appendixdetailedresults}, Table~\ref{tab:detailedsize} (separator size and balance) and Table~\ref{tab:detailedtime} (running times).

We now summarize the results. 
First of all, only using flow-based local search during uncoarsening is already highly competitive, even for small flow problems with $\alpha=0$. 
On average, Flow$_0$ computes 6.7\% smaller separators than Metis and 57\% than Scotch. 
It computes a smaller or equally sized separator than Metis in 89\% of the cases and than Scotch in \emph{every} case. However, it also needs more time to compute a result. This is due to the large flow problems that have to be solved.
Indeed, increasing the value of $\alpha$, \ie searching for separators in larger areas around the initial separator, improves the objective further at the cost of running time. 
For example, increasing $\alpha$ to 0.5 reduces the average size of the computed separator by 3.2\%, but also increases the running time by more than a factor~2 on average. 
Using even larger values of $\alpha>1$ did not further improve the result so that we do not include the data here.
Adding non-flow-based local search also helps to improve the size of the separator. For example, it improves the separator size by 1.8\% when using $\alpha=0$. However, the impact of non-flow-based local search decreases for larger values of $\alpha$.

The strongest configuration of our algorithm is LSFlow$_{1}$. It computes smaller or equally sized separators than Metis in all but two cases and than Scotch in every case. 
On average, separators are~10.3\% smaller than the separators computed by Metis and 62.2\% than the ones computed by 
\begin{figure}[h!]
\centering
\vspace*{-.5cm}
\includegraphics[width=8cm]{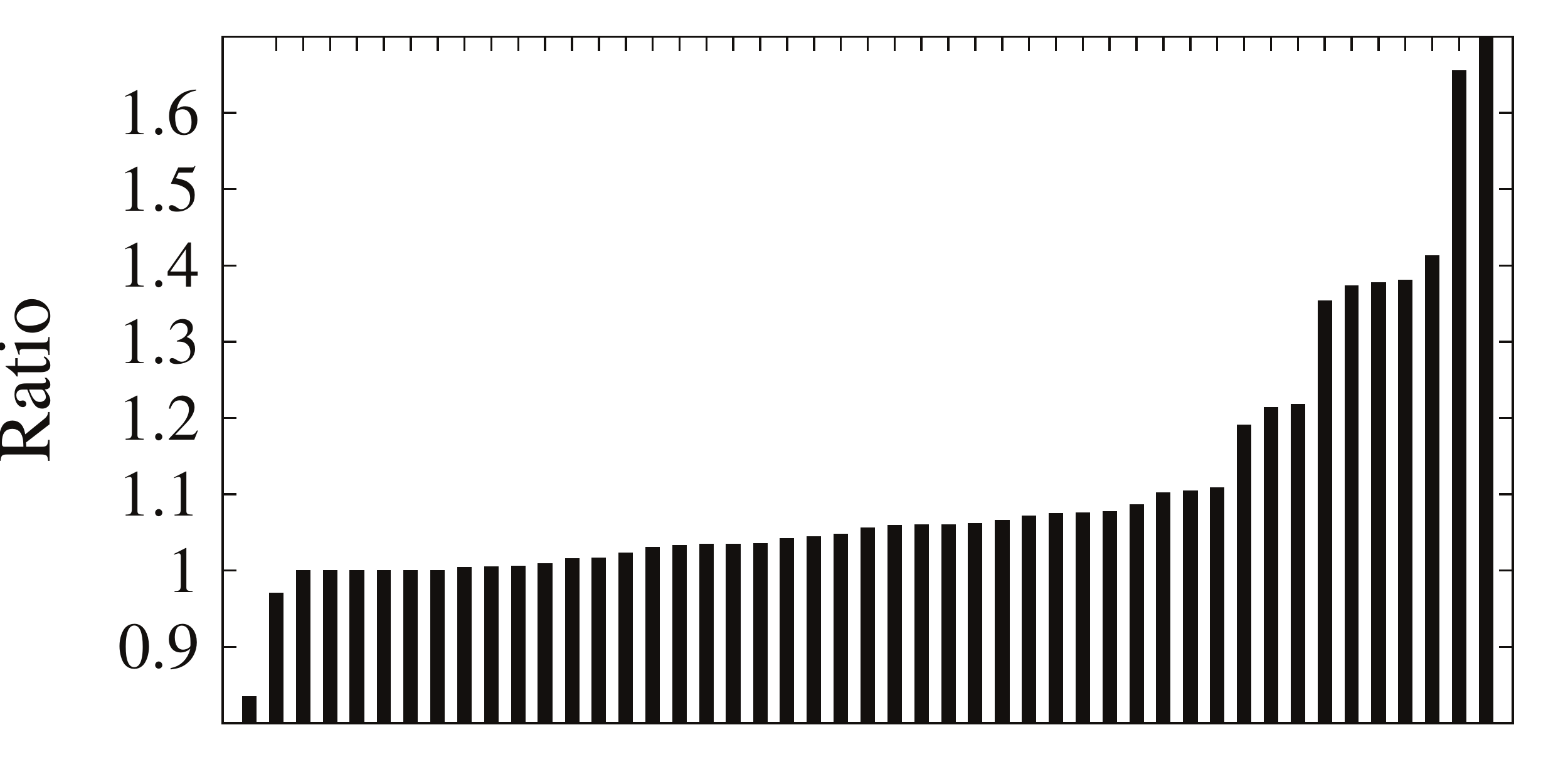} \includegraphics[width=8cm]{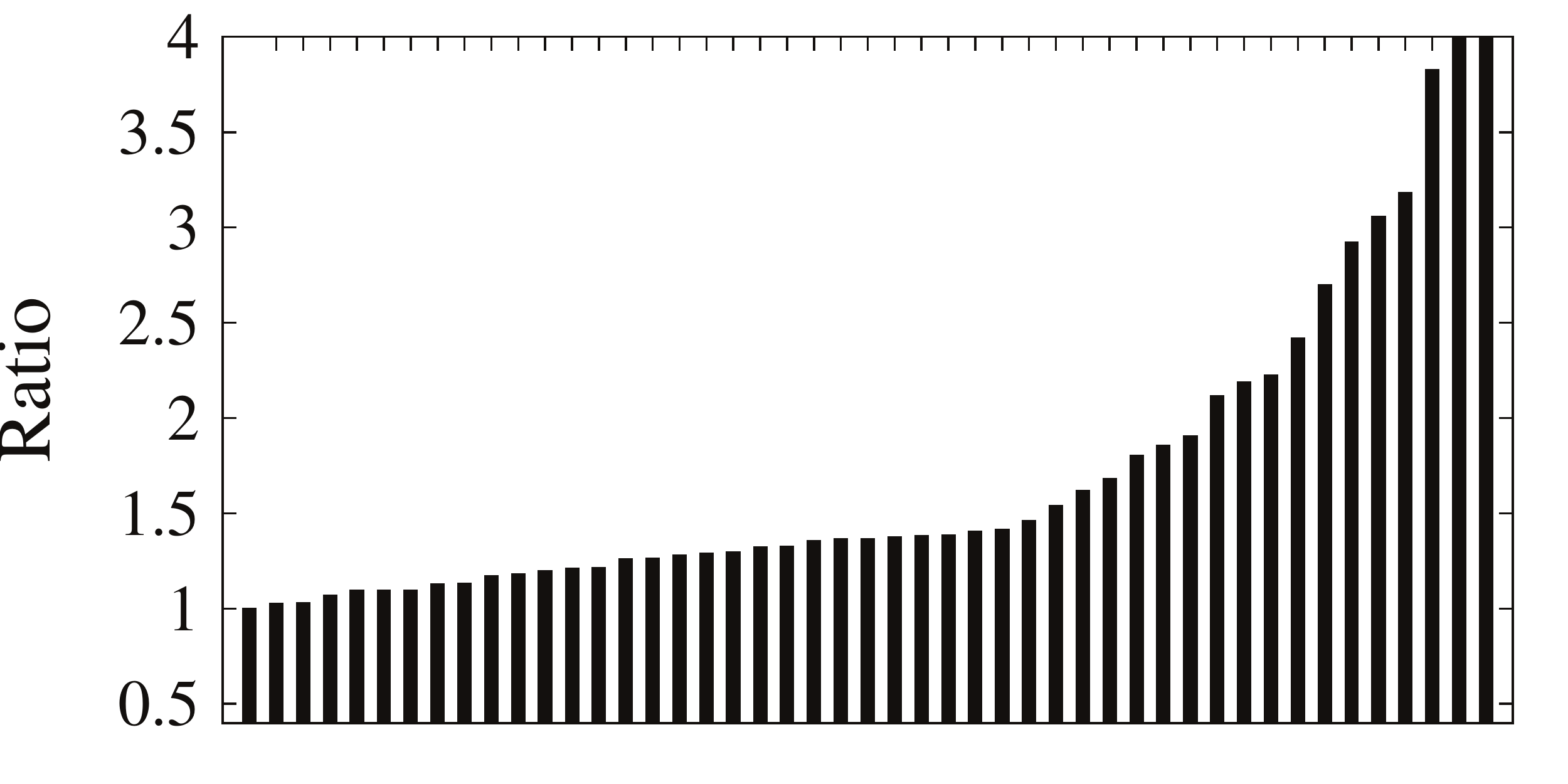}
\caption{Improvement of LSFlow$_{1}$ per instance over Metis (left) and Scotch (right) sorted by absolute value of ratio.}
\label{fig:performanceratios}
\end{figure}
Scotch.
Figure~\ref{fig:performanceratios} shows the average improvement ratios over Metis and Scotch on a per instance basis, sorted by absolute value of improvement. 
The largest improvement over Metis was obtained on the road network europe where our separator is a factor 2.3 smaller whereas the largest improvement over Scotch is on \Id{add32} where our separator is a factor 12 smaller.
On the instance \Id{G2_circuit} Metis computes a 19.9\% smaller separator which is the largest improvement of Metis over our algorithm.

We now compare the size of our separators against the recently published data by LaSalle and Karypis~\cite{lasalle2015efficient}.
The networks used therein that are publicly available are $\Id{auto}$, $\Id{nlr}$, $\Id{del24}$ and $\Id{nlpkkt240}$.
On these graphs our strongest configuration computes separators that are 10.7\%, 10.0\%, 20.1\% and 27.1\% smaller than their best configuration (Greedy+Segmented FM), respectively. 

\section{Conclusion}
\label{s:conclusion}
In this work, we derived algorithms to find small node separators in large graphs. 
We presented a multilevel algorithm that employs novel flow-based local search algorithms and transferred techniques successfully used in the graph partitioning field to the node separator problem. 
This includes the usage of edge ratings tailored to our problem to guide the graph coarsening algorithm as well as highly localized local search and iterated multilevel cycles to improve solution quality even further. 
Experiments indicate that using flow-based local search algorithms as only local search algorithm in a multilevel framework is already highly competitive in terms of separator quality. 

Important future work includes shared-memory parallelization of our algorithms, e.g. currently most of the running time in our algorithm is consumed by the max-flow solver so that a parallel solver will speed up computations. In addition, it is possible to define a simple evolutionary algorithm for the node separator problem by transferring the iterated multilevel scheme to multiple input separators. This will likely result in even better solutions.
\bibliographystyle{plain}
\bibliography{phdthesiscs}
\vfill
\pagebreak
\begin{appendix}
\section{Benchmark Set}
\label{apdx:graphs}
\begin{table}[H]
	\centering
	\begin{tabular}{| l | r | r || l | r | r | }
			\hline
		 	Graph & $n$& $m$ & Graph & $n$& $m$\\
		 	\hline \hline
		 	\multicolumn{3}{|c||}{ Small Walshaw Graphs} &  \multicolumn{3}{c|}{UF Graphs}\\
			\hline

		 	add20        & \numprint{2395}  & \numprint{7462}    & cop20k\_A*                                     & \numprint{99843}  & \numprint{1262244}\\
		 	data         & \numprint{2851}  & \numprint{15093}   & 2cubes\_sphere*                                & \numprint{101492} & \numprint{772886}\\
		 	3elt         & \numprint{4720}  & \numprint{13722}   & thermomech\_TC                                 & \numprint{102158} & \numprint{304700}\\
		 	uk           & \numprint{4824}  & \numprint{6837}    & cfd2                                           & \numprint{123440} & \numprint{1482229}\\
		 	add32        & \numprint{4960}  & \numprint{9462}    & boneS01                                        & \numprint{127224} & \numprint{3293964}\\
		 	bcsstk33     & \numprint{8738}  & \numprint{291583}  & Dubcova3                                       & \numprint{146689} & \numprint{1744980}\\
		 	whitaker3    & \numprint{9800}  & \numprint{28989}   & bmwcra\_1                                      & \numprint{148770} & \numprint{5247616}\\
		 	crack        & \numprint{10240} & \numprint{30380}   & G2\_circuit                                    & \numprint{150102} & \numprint{288286} \\
		 	wing\_nodal* & \numprint{10937} & \numprint{75488}   & c-73                                           & \numprint{169422} & \numprint{554926} \\
		 	fe\_4elt2    & \numprint{11143} & \numprint{32818}   & shipsec5                                       & \numprint{179860} & \numprint{4966618}\\
		 	vibrobox     & \numprint{12328} & \numprint{165250}  & cont-300                                       & \numprint{180895} & \numprint{448799}  \\
		 	\cline{4-6}
		 	bcsstk29*    & \numprint{13992} & \numprint{302748}  & \multicolumn{3}{c|}{ Large Walshaw Graphs}  \\
		 	\cline{4-6}
		 	4elt         & \numprint{15606} & \numprint{45878}   & 598a                                           & \numprint{110971} & \numprint{741934}   \\
		 	fe\_sphere   & \numprint{16386} & \numprint{49152}   & fe\_ocean                                      & \numprint{143437} & \numprint{409593}   \\
		 	cti          & \numprint{16840} & \numprint{48232}   & 144                                            & \numprint{144649} & \numprint{1074393}  \\
		 	memplus      & \numprint{17758} & \numprint{54196}   & wave                                           & \numprint{156317} & \numprint{1059331} \\
		 	cs4          & \numprint{22499} & \numprint{43858}   & m14b                                           & \numprint{214765} & \numprint{1679018}  \\
		 	bcsstk30     & \numprint{28924} & \numprint{1007284} & auto                                           & \numprint{448695} & \numprint{3314611}  \\
		 	\cline{4-6}
		 	bcsstk31     & \numprint{35588} & \numprint{572914}  &  \multicolumn{3}{c|}{ Large Other Graphs}\\
		 	\cline{4-6}
		 	fe\_pwt      & \numprint{36519} & \numprint{144794}  & del23                                          & $\approx$8.4M     & $\approx$25.2M \\
		 	bcsstk32     & \numprint{44609} & \numprint{985046}  & del24                                          & $\approx$16.7M    & $\approx$50.3M \\
		 	\cline{4-6}
		 	fe\_body     & \numprint{45087} & \numprint{163734}  & rgg23                                          & $\approx$8.4M     & $\approx$63.5M \\
		 	t60k*        & \numprint{60005} & \numprint{89440}   & rgg24                                          & $\approx$16.7M    & $\approx$132.6M\\
		 	\cline{4-6}
		 	wing         & \numprint{62032} & \numprint{121544}  & deu                                            & $\approx$4.4M     & $\approx$5.5M \\
		 	brack2       & \numprint{62631} & \numprint{366559}  & eur                                            & $\approx$18.0M    & $\approx$22.2M \\
		 	\cline{4-6}
		 	finan512*    & \numprint{74752} & \numprint{261120}  & af\_shell9                                     & $\approx$504K     & $\approx$8.5M \\
		 	fe\_tooth    & \numprint{78136} & \numprint{452591}  & thermal2                                       & $\approx$1.2M     & $\approx$3.7M   \\
		 	fe\_rotor    & \numprint{99617} & \numprint{662431}  & nlr                                            & $\approx$4.2M     & $\approx$12.5M \\
                        \hline
                        \hline
		 	             &                  &                    & nlpkkt240                                      & $\approx$27.9M    & $\approx$373M  \\

		 	\hline
	\end{tabular}
        \vspace*{.25cm}
 	\caption{Basic properties of the instances used for evaluation.}
 	\label{tab:test_instances_walshaw}
\end{table}

\section{Detailed per Instance Results}
\label{sec:appendixdetailedresults}
\begin{landscape}
\setlength{\tabcolsep}{1ex}
\thispagestyle{empty}
\begin{table}
\tiny
\vspace*{-.25cm}
\hspace*{-1.25cm}
\begin{tabular}{l rrr@{\hskip 13pt}rrr @{\hskip 13pt}rrr @{\hskip 13pt}rrr @{\hskip 13pt}rrr @{\hskip 13pt}rrr @{\hskip 13pt}rrr @{\hskip 13pt}rrr @{\hskip 13pt}rrr @{\hskip 13pt}rrr}
\toprule
         & \multicolumn{3}{c}{Metis} & \multicolumn{3}{c}{Scotch} & \multicolumn{3}{c}{LSFlow$_0$} &  \multicolumn{3}{c}{LSFlow$_{0.5}$} &   \multicolumn{3}{c}{LSFlow$_{1}$} &    \multicolumn{3}{c}{Flow$_{0}$} & \multicolumn{3}{c}{Flow$_{0.5}$} &  \multicolumn{3}{c}{Flow$_{1}$}  \\
 Graph & Avg. & Best& Bal. & Avg. & Best& Bal. & Avg. & Best& Bal. & Avg. & Best& Bal. & Avg. & Best& Bal. & Avg. & Best& Bal. & Avg. & Best& Bal. & Avg. & Best& Bal. \\
\cmidrule(r){1-1} \cmidrule(r){2-4} \cmidrule(r){5-7} \cmidrule(r){8-10} \cmidrule(r){11-13} \cmidrule(r){14-16} \cmidrule(r){17-19} \cmidrule(r){20-22} \cmidrule(r){23-25}
\texttt{\detokenize{144}}            & \numprint{1539} & \numprint{1511} & \numprint{1,13} & \numprint{1639} & \numprint{1602} & \numprint{1,00} & \numprint{1482} & \numprint{1467} & \numprint{1,12} & \numprint{1444} & \numprint{1437} & \numprint{1,19} & \numprint{1445} & \numprint{1439} & \numprint{1,19} & \numprint{1495} & \numprint{1481} & \numprint{1,09} & \numprint{1444} & \numprint{1437} & \numprint{1,20} & \numprint{1446} & \numprint{1437} & \numprint{1,19}\\
\texttt{\detokenize{2cubes_sphere}} & \numprint{1398} & \numprint{1335} & \numprint{1,11} & \numprint{1587} & \numprint{1530} & \numprint{1,00} & \numprint{1265} & \numprint{1245} & \numprint{1,14} & \numprint{1228} & \numprint{1221} & \numprint{1,19} & \numprint{1230} & \numprint{1221} & \numprint{1,19} & \numprint{1274} & \numprint{1266} & \numprint{1,11} & \numprint{1237} & \numprint{1221} & \numprint{1,18} & \numprint{1235} & \numprint{1221} & \numprint{1,18}\\
\texttt{\detokenize{3elt}}           & \numprint{42}   & \numprint{42}   & \numprint{1,09} & \numprint{50}   & \numprint{46}   & \numprint{1,00} & \numprint{42}   & \numprint{42}   & \numprint{1,11} & \numprint{42}   & \numprint{42}   & \numprint{1,11} & \numprint{42}   & \numprint{42}   & \numprint{1,11} & \numprint{42}   & \numprint{42}   & \numprint{1,11} & \numprint{42}   & \numprint{42}   & \numprint{1,11} & \numprint{42}   & \numprint{42}   & \numprint{1,11}\\
\texttt{\detokenize{4elt}}           & \numprint{69}   & \numprint{68}   & \numprint{1,02} & \numprint{82}   & \numprint{73}   & \numprint{1,00} & \numprint{68}   & \numprint{68}   & \numprint{1,01} & \numprint{68}   & \numprint{68}   & \numprint{1,01} & \numprint{68}   & \numprint{68}   & \numprint{1,01} & \numprint{68}   & \numprint{68}   & \numprint{1,02} & \numprint{68}   & \numprint{68}   & \numprint{1,01} & \numprint{68}   & \numprint{68}   & \numprint{1,01}\\
\texttt{\detokenize{598a}}           & \numprint{615}  & \numprint{603}  & \numprint{1,03} & \numprint{639}  & \numprint{629}  & \numprint{1,00} & \numprint{594}  & \numprint{593}  & \numprint{1,04} & \numprint{593}  & \numprint{593}  & \numprint{1,03} & \numprint{593}  & \numprint{593}  & \numprint{1,03} & \numprint{594}  & \numprint{593}  & \numprint{1,04} & \numprint{593}  & \numprint{593}  & \numprint{1,03} & \numprint{593}  & \numprint{593}  & \numprint{1,03}\\
\texttt{\detokenize{add20}}          & \numprint{25}   & \numprint{23}   & \numprint{1,09} & \numprint{142}  & \numprint{128}  & \numprint{1,10} & \numprint{26}   & \numprint{23}   & \numprint{1,11} & \numprint{23}   & \numprint{23}   & \numprint{1,08} & \numprint{24}   & \numprint{23}   & \numprint{1,08} & \numprint{28}   & \numprint{23}   & \numprint{1,10} & \numprint{23}   & \numprint{23}   & \numprint{1,08} & \numprint{24}   & \numprint{23}   & \numprint{1,08}\\
\texttt{\detokenize{add32}}          & \numprint{1}    & \numprint{1}    & \numprint{1,08} & \numprint{14}   & \numprint{4}    & \numprint{1,00} & \numprint{1}    & \numprint{1}    & \numprint{1,12} & \numprint{1}    & \numprint{1}    & \numprint{1,12} & \numprint{1}    & \numprint{1}    & \numprint{1,12} & \numprint{1}    & \numprint{1}    & \numprint{1,12} & \numprint{1}    & \numprint{1}    & \numprint{1,12} & \numprint{1}    & \numprint{1}    & \numprint{1,12}\\
\texttt{\detokenize{af_shell9}}     & \numprint{934}  & \numprint{885}  & \numprint{1,00} & \numprint{1382} & \numprint{1095} & \numprint{1,00} & \numprint{880}  & \numprint{880}  & \numprint{1,06} & \numprint{880}  & \numprint{880}  & \numprint{1,06} & \numprint{880}  & \numprint{880}  & \numprint{1,06} & \numprint{880}  & \numprint{880}  & \numprint{1,06} & \numprint{880}  & \numprint{880}  & \numprint{1,06} & \numprint{880}  & \numprint{880}  & \numprint{1,06}\\
\texttt{\detokenize{auto}}           & \numprint{2109} & \numprint{2073} & \numprint{1,18} & \numprint{3158} & \numprint{2547} & \numprint{1,00} & \numprint{2034} & \numprint{2021} & \numprint{1,19} & \numprint{1986} & \numprint{1977} & \numprint{1,20} & \numprint{1992} & \numprint{1978} & \numprint{1,20} & \numprint{2093} & \numprint{2062} & \numprint{1,17} & \numprint{1992} & \numprint{1981} & \numprint{1,20} & \numprint{1988} & \numprint{1978} & \numprint{1,20}\\
\texttt{\detokenize{bcsstk29}}       & \numprint{180}  & \numprint{180}  & \numprint{1,00} & \numprint{260}  & \numprint{234}  & \numprint{1,01} & \numprint{180}  & \numprint{180}  & \numprint{1,02} & \numprint{180}  & \numprint{180}  & \numprint{1,11} & \numprint{180}  & \numprint{180}  & \numprint{1,11} & \numprint{180}  & \numprint{180}  & \numprint{1,01} & \numprint{180}  & \numprint{180}  & \numprint{1,11} & \numprint{180}  & \numprint{180}  & \numprint{1,10}\\
\texttt{\detokenize{bcsstk30}}       & \numprint{208}  & \numprint{206}  & \numprint{1,04} & \numprint{439}  & \numprint{393}  & \numprint{1,02} & \numprint{206}  & \numprint{206}  & \numprint{1,00} & \numprint{206}  & \numprint{206}  & \numprint{1,00} & \numprint{206}  & \numprint{206}  & \numprint{1,00} & \numprint{206}  & \numprint{206}  & \numprint{1,00} & \numprint{206}  & \numprint{206}  & \numprint{1,00} & \numprint{206}  & \numprint{206}  & \numprint{1,00}\\
\texttt{\detokenize{bcsstk31}}       & \numprint{298}  & \numprint{285}  & \numprint{1,07} & \numprint{482}  & \numprint{437}  & \numprint{1,04} & \numprint{271}  & \numprint{268}  & \numprint{1,10} & \numprint{268}  & \numprint{268}  & \numprint{1,17} & \numprint{268}  & \numprint{268}  & \numprint{1,17} & \numprint{271}  & \numprint{270}  & \numprint{1,09} & \numprint{268}  & \numprint{268}  & \numprint{1,17} & \numprint{268}  & \numprint{268}  & \numprint{1,17}\\
\texttt{\detokenize{bcsstk32}}       & \numprint{276}  & \numprint{252}  & \numprint{1,19} & \numprint{752}  & \numprint{463}  & \numprint{1,04} & \numprint{236}  & \numprint{229}  & \numprint{1,19} & \numprint{239}  & \numprint{229}  & \numprint{1,18} & \numprint{232}  & \numprint{229}  & \numprint{1,20} & \numprint{252}  & \numprint{239}  & \numprint{1,17} & \numprint{239}  & \numprint{229}  & \numprint{1,18} & \numprint{233}  & \numprint{229}  & \numprint{1,20}\\
\texttt{\detokenize{bcsstk33}}       & \numprint{421}  & \numprint{421}  & \numprint{0,96} & \numprint{549}  & \numprint{179}  & \numprint{1,21} & \numprint{282}  & \numprint{262}  & \numprint{1,18} & \numprint{267}  & \numprint{261}  & \numprint{1,20} & \numprint{283}  & \numprint{265}  & \numprint{1,19} & \numprint{292}  & \numprint{274}  & \numprint{1,17} & \numprint{272}  & \numprint{266}  & \numprint{1,20} & \numprint{288}  & \numprint{266}  & \numprint{1,19}\\
\texttt{\detokenize{bmwcra_1}}      & \numprint{318}  & \numprint{318}  & \numprint{1,13} & \numprint{1006} & \numprint{576}  & \numprint{1,06} & \numprint{318}  & \numprint{318}  & \numprint{1,14} & \numprint{350}  & \numprint{318}  & \numprint{1,13} & \numprint{350}  & \numprint{318}  & \numprint{1,13} & \numprint{318}  & \numprint{318}  & \numprint{1,14} & \numprint{350}  & \numprint{318}  & \numprint{1,13} & \numprint{350}  & \numprint{318}  & \numprint{1,13}\\
\texttt{\detokenize{boneS01}}        & \numprint{1583} & \numprint{1542} & \numprint{1,08} & \numprint{4137} & \numprint{3969} & \numprint{1,00} & \numprint{1525} & \numprint{1500} & \numprint{1,04} & \numprint{1500} & \numprint{1500} & \numprint{1,10} & \numprint{1500} & \numprint{1500} & \numprint{1,10} & \numprint{1524} & \numprint{1500} & \numprint{1,04} & \numprint{1500} & \numprint{1500} & \numprint{1,10} & \numprint{1500} & \numprint{1500} & \numprint{1,10}\\
\texttt{\detokenize{brack2}}         & \numprint{182}  & \numprint{181}  & \numprint{1,07} & \numprint{237}  & \numprint{214}  & \numprint{1,00} & \numprint{181}  & \numprint{181}  & \numprint{1,07} & \numprint{181}  & \numprint{181}  & \numprint{1,07} & \numprint{181}  & \numprint{181}  & \numprint{1,07} & \numprint{181}  & \numprint{181}  & \numprint{1,07} & \numprint{181}  & \numprint{181}  & \numprint{1,07} & \numprint{181}  & \numprint{181}  & \numprint{1,07}\\
\texttt{\detokenize{cfd2}}           & \numprint{1040} & \numprint{1030} & \numprint{1,05} & \numprint{1303} & \numprint{1163} & \numprint{1,00} & \numprint{1030} & \numprint{1030} & \numprint{1,06} & \numprint{1030} & \numprint{1030} & \numprint{1,09} & \numprint{1030} & \numprint{1030} & \numprint{1,08} & \numprint{1030} & \numprint{1030} & \numprint{1,06} & \numprint{1030} & \numprint{1030} & \numprint{1,08} & \numprint{1030} & \numprint{1030} & \numprint{1,07}\\
\texttt{\detokenize{cont-300}}       & \numprint{598}  & \numprint{598}  & \numprint{1,00} & \numprint{616}  & \numprint{598}  & \numprint{1,00} & \numprint{598}  & \numprint{598}  & \numprint{1,00} & \numprint{598}  & \numprint{598}  & \numprint{1,00} & \numprint{579}  & \numprint{534}  & \numprint{1,06} & \numprint{598}  & \numprint{598}  & \numprint{1,02} & \numprint{598}  & \numprint{598}  & \numprint{1,18} & \numprint{598}  & \numprint{598}  & \numprint{1,18}\\
\texttt{\detokenize{cop20k_A}}      & \numprint{680}  & \numprint{660}  & \numprint{1,02} & \numprint{1904} & \numprint{1833} & \numprint{1,00} & \numprint{613}  & \numprint{613}  & \numprint{1,04} & \numprint{613}  & \numprint{613}  & \numprint{1,04} & \numprint{613}  & \numprint{613}  & \numprint{1,04} & \numprint{613}  & \numprint{613}  & \numprint{1,04} & \numprint{613}  & \numprint{613}  & \numprint{1,04} & \numprint{613}  & \numprint{613}  & \numprint{1,04}\\
\texttt{\detokenize{crack}}          & \numprint{72}   & \numprint{69}   & \numprint{1,08} & \numprint{92}   & \numprint{81}   & \numprint{1,00} & \numprint{69}   & \numprint{68}   & \numprint{1,13} & \numprint{68}   & \numprint{68}   & \numprint{1,16} & \numprint{68}   & \numprint{68}   & \numprint{1,16} & \numprint{69}   & \numprint{68}   & \numprint{1,13} & \numprint{68}   & \numprint{68}   & \numprint{1,16} & \numprint{68}   & \numprint{68}   & \numprint{1,16}\\
\texttt{\detokenize{cs4}}            & \numprint{289}  & \numprint{281}  & \numprint{1,11} & \numprint{332}  & \numprint{323}  & \numprint{1,00} & \numprint{281}  & \numprint{279}  & \numprint{1,09} & \numprint{267}  & \numprint{264}  & \numprint{1,19} & \numprint{268}  & \numprint{264}  & \numprint{1,19} & \numprint{284}  & \numprint{282}  & \numprint{1,08} & \numprint{267}  & \numprint{265}  & \numprint{1,19} & \numprint{269}  & \numprint{265}  & \numprint{1,18}\\
\texttt{\detokenize{cti}}            & \numprint{268}  & \numprint{266}  & \numprint{1,00} & \numprint{291}  & \numprint{283}  & \numprint{1,00} & \numprint{267}  & \numprint{266}  & \numprint{0,99} & \numprint{266}  & \numprint{266}  & \numprint{0,98} & \numprint{266}  & \numprint{266}  & \numprint{0,98} & \numprint{267}  & \numprint{266}  & \numprint{1,01} & \numprint{266}  & \numprint{266}  & \numprint{1,00} & \numprint{266}  & \numprint{266}  & \numprint{1,00}\\
\texttt{\detokenize{data}}           & \numprint{59}   & \numprint{45}   & \numprint{1,10} & \numprint{69}   & \numprint{64}   & \numprint{1,00} & \numprint{44}   & \numprint{41}   & \numprint{1,17} & \numprint{42}   & \numprint{41}   & \numprint{1,18} & \numprint{43}   & \numprint{41}   & \numprint{1,18} & \numprint{45}   & \numprint{43}   & \numprint{1,15} & \numprint{42}   & \numprint{41}   & \numprint{1,17} & \numprint{43}   & \numprint{41}   & \numprint{1,18}\\
\texttt{\detokenize{del23}}          & \numprint{2486} & \numprint{2434} & \numprint{1,03} & \numprint{2933} & \numprint{2741} & \numprint{1,00} & \numprint{2050} & \numprint{2048} & \numprint{1,01} & \numprint{2048} & \numprint{2048} & \numprint{1,05} & \numprint{2048} & \numprint{2048} & \numprint{1,04} & \numprint{2050} & \numprint{2048} & \numprint{1,01} & \numprint{2048} & \numprint{2048} & \numprint{1,04} & \numprint{2048} & \numprint{2048} & \numprint{1,04}\\
\texttt{\detokenize{del24}}          & \numprint{3541} & \numprint{3472} & \numprint{1,01} & \numprint{4004} & \numprint{3792} & \numprint{1,00} & \numprint{2908} & \numprint{2904} & \numprint{1,01} & \numprint{2907} & \numprint{2904} & \numprint{1,03} & \numprint{2907} & \numprint{2904} & \numprint{1,03} & \numprint{2908} & \numprint{2904} & \numprint{1,01} & \numprint{2907} & \numprint{2904} & \numprint{1,03} & \numprint{2907} & \numprint{2904} & \numprint{1,03}\\
\texttt{\detokenize{deu}}            & \numprint{241}  & \numprint{217}  & \numprint{1,07} & \numprint{325}  & \numprint{286}  & \numprint{1,00} & \numprint{152}  & \numprint{152}  & \numprint{1,04} & \numprint{145}  & \numprint{145}  & \numprint{1,12} & \numprint{145}  & \numprint{145}  & \numprint{1,12} & \numprint{152}  & \numprint{152}  & \numprint{1,04} & \numprint{145}  & \numprint{145}  & \numprint{1,12} & \numprint{145}  & \numprint{145}  & \numprint{1,12}\\
\texttt{\detokenize{Dubcova3}}       & \numprint{406}  & \numprint{383}  & \numprint{1,02} & \numprint{1495} & \numprint{1395} & \numprint{1,00} & \numprint{383}  & \numprint{383}  & \numprint{1,04} & \numprint{383}  & \numprint{383}  & \numprint{1,16} & \numprint{383}  & \numprint{383}  & \numprint{1,15} & \numprint{383}  & \numprint{383}  & \numprint{1,05} & \numprint{383}  & \numprint{383}  & \numprint{1,16} & \numprint{383}  & \numprint{383}  & \numprint{1,18}\\
\texttt{\detokenize{eur}}            & \numprint{430}  & \numprint{349}  & \numprint{1,09} & \numprint{620}  & \numprint{486}  & \numprint{1,01} & \numprint{218}  & \numprint{109}  & \numprint{1,07} & \numprint{208}  & \numprint{200}  & \numprint{1,12} & \numprint{206}  & \numprint{195}  & \numprint{1,13} & \numprint{218}  & \numprint{109}  & \numprint{1,07} & \numprint{208}  & \numprint{200}  & \numprint{1,12} & \numprint{206}  & \numprint{195}  & \numprint{1,13}\\
\texttt{\detokenize{fe_4elt2}}      & \numprint{66}   & \numprint{66}   & \numprint{0,99} & \numprint{69}   & \numprint{67}   & \numprint{1,00} & \numprint{66}   & \numprint{66}   & \numprint{0,99} & \numprint{66}   & \numprint{66}   & \numprint{0,99} & \numprint{66}   & \numprint{66}   & \numprint{0,99} & \numprint{66}   & \numprint{66}   & \numprint{1,02} & \numprint{66}   & \numprint{66}   & \numprint{1,04} & \numprint{66}   & \numprint{66}   & \numprint{1,04}\\
\texttt{\detokenize{fe_body}}       & \numprint{86}   & \numprint{65}   & \numprint{1,11} & \numprint{160}  & \numprint{122}  & \numprint{1,01} & \numprint{78}   & \numprint{66}   & \numprint{1,12} & \numprint{77}   & \numprint{61}   & \numprint{1,15} & \numprint{75}   & \numprint{62}   & \numprint{1,14} & \numprint{78}   & \numprint{66}   & \numprint{1,12} & \numprint{77}   & \numprint{61}   & \numprint{1,15} & \numprint{75}   & \numprint{62}   & \numprint{1,14}\\
\texttt{\detokenize{fe_ocean}}      & \numprint{273}  & \numprint{263}  & \numprint{1,01} & \numprint{340}  & \numprint{322}  & \numprint{1,00} & \numprint{263}  & \numprint{263}  & \numprint{1,02} & \numprint{263}  & \numprint{263}  & \numprint{1,02} & \numprint{263}  & \numprint{263}  & \numprint{1,02} & \numprint{263}  & \numprint{263}  & \numprint{1,02} & \numprint{263}  & \numprint{263}  & \numprint{1,02} & \numprint{263}  & \numprint{263}  & \numprint{1,02}\\
\texttt{\detokenize{fe_pwt}}        & \numprint{120}  & \numprint{120}  & \numprint{1,01} & \numprint{132}  & \numprint{124}  & \numprint{1,00} & \numprint{116}  & \numprint{116}  & \numprint{1,03} & \numprint{116}  & \numprint{116}  & \numprint{1,09} & \numprint{116}  & \numprint{116}  & \numprint{1,12} & \numprint{116}  & \numprint{116}  & \numprint{1,03} & \numprint{116}  & \numprint{116}  & \numprint{1,13} & \numprint{116}  & \numprint{116}  & \numprint{1,13}\\
\texttt{\detokenize{fe_rotor}}      & \numprint{453}  & \numprint{441}  & \numprint{1,04} & \numprint{576}  & \numprint{514}  & \numprint{1,05} & \numprint{441}  & \numprint{439}  & \numprint{1,07} & \numprint{441}  & \numprint{439}  & \numprint{1,08} & \numprint{441}  & \numprint{439}  & \numprint{1,07} & \numprint{441}  & \numprint{439}  & \numprint{1,08} & \numprint{442}  & \numprint{439}  & \numprint{1,08} & \numprint{442}  & \numprint{439}  & \numprint{1,08}\\
\texttt{\detokenize{fe_sphere}}     & \numprint{195}  & \numprint{192}  & \numprint{0,99} & \numprint{239}  & \numprint{227}  & \numprint{1,00} & \numprint{192}  & \numprint{192}  & \numprint{1,04} & \numprint{192}  & \numprint{192}  & \numprint{1,05} & \numprint{192}  & \numprint{192}  & \numprint{1,05} & \numprint{192}  & \numprint{192}  & \numprint{1,02} & \numprint{192}  & \numprint{192}  & \numprint{1,13} & \numprint{192}  & \numprint{192}  & \numprint{1,14}\\
\texttt{\detokenize{fe_tooth}}      & \numprint{882}  & \numprint{867}  & \numprint{1,16} & \numprint{1192} & \numprint{1094} & \numprint{1,00} & \numprint{882}  & \numprint{869}  & \numprint{1,13} & \numprint{849}  & \numprint{837}  & \numprint{1,19} & \numprint{848}  & \numprint{826}  & \numprint{1,19} & \numprint{885}  & \numprint{882}  & \numprint{1,11} & \numprint{852}  & \numprint{827}  & \numprint{1,19} & \numprint{853}  & \numprint{839}  & \numprint{1,19}\\
\texttt{\detokenize{finan512}}       & \numprint{50}   & \numprint{50}   & \numprint{1,07} & \numprint{67}   & \numprint{51}   & \numprint{1,02} & \numprint{50}   & \numprint{50}   & \numprint{1,01} & \numprint{50}   & \numprint{50}   & \numprint{1,13} & \numprint{50}   & \numprint{50}   & \numprint{1,13} & \numprint{50}   & \numprint{50}   & \numprint{1,01} & \numprint{50}   & \numprint{50}   & \numprint{1,12} & \numprint{50}   & \numprint{50}   & \numprint{1,13}\\
\texttt{\detokenize{G2_circuit}}    & \numprint{312}  & \numprint{312}  & \numprint{1,03} & \numprint{416}  & \numprint{348}  & \numprint{1,00} & \numprint{374}  & \numprint{312}  & \numprint{1,01} & \numprint{374}  & \numprint{312}  & \numprint{1,03} & \numprint{374}  & \numprint{312}  & \numprint{1,03} & \numprint{374}  & \numprint{312}  & \numprint{1,02} & \numprint{374}  & \numprint{312}  & \numprint{1,14} & \numprint{374}  & \numprint{312}  & \numprint{1,14}\\
\texttt{\detokenize{m14b}}           & \numprint{885}  & \numprint{859}  & \numprint{1,04} & \numprint{895}  & \numprint{870}  & \numprint{1,00} & \numprint{835}  & \numprint{834}  & \numprint{1,02} & \numprint{834}  & \numprint{834}  & \numprint{1,00} & \numprint{834}  & \numprint{834}  & \numprint{1,00} & \numprint{835}  & \numprint{834}  & \numprint{1,02} & \numprint{834}  & \numprint{834}  & \numprint{1,00} & \numprint{834}  & \numprint{834}  & \numprint{1,00}\\
\texttt{\detokenize{memplus}}        & \numprint{88}   & \numprint{81}   & \numprint{1,19} & \numprint{95}   & \numprint{95}   & \numprint{1,00} & \numprint{81}   & \numprint{72}   & \numprint{1,15} & \numprint{66}   & \numprint{62}   & \numprint{1,15} & \numprint{68}   & \numprint{65}   & \numprint{1,15} & \numprint{108}  & \numprint{76}   & \numprint{1,10} & \numprint{70}   & \numprint{65}   & \numprint{1,12} & \numprint{72}   & \numprint{68}   & \numprint{1,11}\\
\texttt{\detokenize{nlr}}            & \numprint{1823} & \numprint{1805} & \numprint{1,01} & \numprint{2156} & \numprint{1991} & \numprint{1,00} & \numprint{1663} & \numprint{1663} & \numprint{1,04} & \numprint{1655} & \numprint{1655} & \numprint{1,17} & \numprint{1655} & \numprint{1655} & \numprint{1,17} & \numprint{1663} & \numprint{1663} & \numprint{1,04} & \numprint{1655} & \numprint{1655} & \numprint{1,17} & \numprint{1655} & \numprint{1655} & \numprint{1,17}\\
\texttt{\detokenize{rgg23}}          & \numprint{3395} & \numprint{3327} & \numprint{1,02} & \numprint{3466} & \numprint{3298} & \numprint{1,00} & \numprint{2475} & \numprint{2471} & \numprint{1,09} & \numprint{2473} & \numprint{2470} & \numprint{1,14} & \numprint{2473} & \numprint{2470} & \numprint{1,14} & \numprint{2475} & \numprint{2471} & \numprint{1,09} & \numprint{2473} & \numprint{2470} & \numprint{1,14} & \numprint{2473} & \numprint{2470} & \numprint{1,14}\\
\texttt{\detokenize{rgg24}}          & \numprint{5020} & \numprint{4850} & \numprint{1,02} & \numprint{5073} & \numprint{4961} & \numprint{1,00} & \numprint{3648} & \numprint{3636} & \numprint{1,13} & \numprint{3644} & \numprint{3636} & \numprint{1,14} & \numprint{3644} & \numprint{3636} & \numprint{1,14} & \numprint{3648} & \numprint{3636} & \numprint{1,13} & \numprint{3644} & \numprint{3636} & \numprint{1,14} & \numprint{3644} & \numprint{3636} & \numprint{1,14}\\
\texttt{\detokenize{shipsec5}}       & \numprint{1222} & \numprint{1191} & \numprint{1,05} & \numprint{2031} & \numprint{1887} & \numprint{1,00} & \numprint{1199} & \numprint{1191} & \numprint{1,02} & \numprint{1185} & \numprint{1185} & \numprint{1,16} & \numprint{1185} & \numprint{1185} & \numprint{1,16} & \numprint{1202} & \numprint{1191} & \numprint{1,00} & \numprint{1185} & \numprint{1185} & \numprint{1,16} & \numprint{1185} & \numprint{1185} & \numprint{1,16}\\
\texttt{\detokenize{t60k}}           & \numprint{58}   & \numprint{56}   & \numprint{1,09} & \numprint{97}   & \numprint{87}   & \numprint{1,00} & \numprint{56}   & \numprint{56}   & \numprint{1,10} & \numprint{56}   & \numprint{56}   & \numprint{1,10} & \numprint{56}   & \numprint{56}   & \numprint{1,10} & \numprint{56}   & \numprint{56}   & \numprint{1,10} & \numprint{56}   & \numprint{56}   & \numprint{1,10} & \numprint{56}   & \numprint{56}   & \numprint{1,10}\\
\texttt{\detokenize{thermal2}}       & \numprint{468}  & \numprint{462}  & \numprint{1,02} & \numprint{524}  & \numprint{494}  & \numprint{1,00} & \numprint{430}  & \numprint{430}  & \numprint{1,03} & \numprint{430}  & \numprint{430}  & \numprint{1,03} & \numprint{430}  & \numprint{430}  & \numprint{1,03} & \numprint{430}  & \numprint{430}  & \numprint{1,03} & \numprint{430}  & \numprint{430}  & \numprint{1,03} & \numprint{430}  & \numprint{430}  & \numprint{1,03}\\
\texttt{\detokenize{thermomech}}     & \numprint{132}  & \numprint{129}  & \numprint{1,03} & \numprint{153}  & \numprint{136}  & \numprint{1,00} & \numprint{126}  & \numprint{126}  & \numprint{1,06} & \numprint{126}  & \numprint{126}  & \numprint{1,07} & \numprint{126}  & \numprint{126}  & \numprint{1,07} & \numprint{126}  & \numprint{126}  & \numprint{1,06} & \numprint{126}  & \numprint{126}  & \numprint{1,07} & \numprint{126}  & \numprint{126}  & \numprint{1,07}\\
\texttt{\detokenize{uk}}             & \numprint{15}   & \numprint{14}   & \numprint{1,16} & \numprint{25}   & \numprint{21}   & \numprint{1,00} & \numprint{14}   & \numprint{14}   & \numprint{1,19} & \numprint{14}   & \numprint{14}   & \numprint{1,19} & \numprint{14}   & \numprint{14}   & \numprint{1,19} & \numprint{15}   & \numprint{14}   & \numprint{1,18} & \numprint{14}   & \numprint{14}   & \numprint{1,19} & \numprint{14}   & \numprint{14}   & \numprint{1,19}\\
\texttt{\detokenize{vibrobox}}       & \numprint{582}  & \numprint{554}  & \numprint{1,14} & \numprint{967}  & \numprint{756}  & \numprint{0,92} & \numprint{643}  & \numprint{554}  & \numprint{1,12} & \numprint{581}  & \numprint{554}  & \numprint{1,16} & \numprint{614}  & \numprint{554}  & \numprint{1,13} & \numprint{826}  & \numprint{598}  & \numprint{1,07} & \numprint{581}  & \numprint{554}  & \numprint{1,16} & \numprint{614}  & \numprint{554}  & \numprint{1,12}\\
\texttt{\detokenize{wave}}           & \numprint{2254} & \numprint{2204} & \numprint{1,02} & \numprint{2451} & \numprint{2329} & \numprint{1,00} & \numprint{2168} & \numprint{2122} & \numprint{1,07} & \numprint{2114} & \numprint{2079} & \numprint{1,15} & \numprint{2101} & \numprint{2077} & \numprint{1,16} & \numprint{2174} & \numprint{2112} & \numprint{1,06} & \numprint{2121} & \numprint{2080} & \numprint{1,15} & \numprint{2101} & \numprint{2079} & \numprint{1,17}\\
\texttt{\detokenize{whitaker3}}      & \numprint{64}   & \numprint{63}   & \numprint{1,02} & \numprint{70}   & \numprint{67}   & \numprint{1,00} & \numprint{63}   & \numprint{63}   & \numprint{0,99} & \numprint{62}   & \numprint{62}   & \numprint{1,19} & \numprint{62}   & \numprint{62}   & \numprint{1,19} & \numprint{63}   & \numprint{63}   & \numprint{1,00} & \numprint{62}   & \numprint{62}   & \numprint{1,19} & \numprint{62}   & \numprint{62}   & \numprint{1,19}\\
\texttt{\detokenize{wing}}           & \numprint{630}  & \numprint{607}  & \numprint{1,12} & \numprint{612}  & \numprint{188}  & \numprint{1,16} & \numprint{613}  & \numprint{605}  & \numprint{1,10} & \numprint{590}  & \numprint{583}  & \numprint{1,19} & \numprint{586}  & \numprint{584}  & \numprint{1,18} & \numprint{615}  & \numprint{608}  & \numprint{1,08} & \numprint{589}  & \numprint{581}  & \numprint{1,18} & \numprint{587}  & \numprint{583}  & \numprint{1,18}\\
\texttt{\detokenize{wing_nodal}}    & \numprint{389}  & \numprint{383}  & \numprint{1,17} & \numprint{381}  & \numprint{167}  & \numprint{1,23} & \numprint{386}  & \numprint{378}  & \numprint{1,16} & \numprint{375}  & \numprint{374}  & \numprint{1,20} & \numprint{375}  & \numprint{374}  & \numprint{1,19} & \numprint{407}  & \numprint{406}  & \numprint{1,06} & \numprint{375}  & \numprint{374}  & \numprint{1,19} & \numprint{375}  & \numprint{374}  & \numprint{1,19}\\
\bottomrule
              \end{tabular}
              \vspace*{.5cm}
\caption{Detailed per instances results as average and best values for the size of separator and average balance.}                
\label{tab:detailedsize}
\end{table}
\end{landscape}
\setlength{\tabcolsep}{1ex}
\begin{table}[h!]
\scriptsize
\centering
\begin{tabular}{l rr rrr rrr}
\toprule
                        & {Metis}         & {Scotch}        & {LSFlow$_0$}      & {LSFlow$_{0.5}$}   & {LSFlow$_{1}$}     & {Flow$_{0}$}      & {Flow$_{0.5}$}     & {Flow$_{1}$}       \\
Graph          & $t_\text{avg.}$ & $t_\text{avg.}$ & $t_\text{avg.}$   & $t_\text{avg.}$    & $t_\text{avg.}$    & $t_\text{avg.}$   & $t_\text{avg.}$    & $t_\text{avg.}$    \\
\midrule
\texttt{\detokenize{144}}            & \numprint{0,2}  & \numprint{0,3} & \numprint{85,6}   & \numprint{132,6}   & \numprint{166,3}   & \numprint{27,0}   & \numprint{82,8}    & \numprint{95,9}\\
\texttt{\detokenize{2cubes_sphere}} & \numprint{0,1}  & \numprint{0,2} & \numprint{67,5}   & \numprint{106,4}   & \numprint{124,8}   & \numprint{21,5}   & \numprint{62,6}    & \numprint{82,9}\\
\texttt{\detokenize{3elt}}           & \numprint{0,1}  & \numprint{0,1} & \numprint{1,2}    & \numprint{1,3}     & \numprint{1,5}     & \numprint{1,0}    & \numprint{1,2}     & \numprint{1,3}\\
\texttt{\detokenize{4elt}}           & \numprint{0,1}  & \numprint{0,1} & \numprint{2,1}    & \numprint{2,9}     & \numprint{3,5}     & \numprint{1,6}    & \numprint{2,6}     & \numprint{3,1}\\
\texttt{\detokenize{598a}}           & \numprint{0,1}  & \numprint{0,2} & \numprint{31,5}   & \numprint{48,4}    & \numprint{59,5}    & \numprint{14,2}   & \numprint{32,8}    & \numprint{44,1}\\
\texttt{\detokenize{add20}}          & \numprint{0,1}  & \numprint{0,1} & \numprint{6,1}    & \numprint{5,8}     & \numprint{5,0}     & \numprint{5,4}    & \numprint{5,3}     & \numprint{4,4}\\
\texttt{\detokenize{add32}}          & \numprint{0,1}  & \numprint{0,1} & \numprint{0,8}    & \numprint{0,8}     & \numprint{0,9}     & \numprint{0,7}    & \numprint{0,8}     & \numprint{0,9}\\
\texttt{\detokenize{af_shell9}}     & \numprint{0,6}  & \numprint{1,3} & \numprint{140,1}  & \numprint{278,6}   & \numprint{343,8}   & \numprint{103,6}  & \numprint{237,9}   & \numprint{339,6}\\
\texttt{\detokenize{auto}}           & \numprint{0,6}  & \numprint{1,0} & \numprint{146,0}  & \numprint{468,7}   & \numprint{603,9}   & \numprint{65,3}   & \numprint{386,2}   & \numprint{450,1}\\
\texttt{\detokenize{bcsstk29}}       & \numprint{0,1}  & \numprint{0,4} & \numprint{8,1}    & \numprint{9,4}     & \numprint{10,2}    & \numprint{5,4}    & \numprint{7,0}     & \numprint{7,9}\\
\texttt{\detokenize{bcsstk30}}       & \numprint{0,1}  & \numprint{1,2} & \numprint{26,7}   & \numprint{38,6}    & \numprint{42,9}    & \numprint{11,8}   & \numprint{23,1}    & \numprint{29,2}\\
\texttt{\detokenize{bcsstk31}}       & \numprint{0,1}  & \numprint{0,4} & \numprint{17,6}   & \numprint{24,0}    & \numprint{25,8}    & \numprint{7,6}    & \numprint{12,1}    & \numprint{15,1}\\
\texttt{\detokenize{bcsstk32}}       & \numprint{0,1}  & \numprint{0,9} & \numprint{21,0}   & \numprint{35,2}    & \numprint{39,7}    & \numprint{9,6}    & \numprint{24,0}    & \numprint{32,5}\\
\texttt{\detokenize{bcsstk33}}       & \numprint{0,1}  & \numprint{1,4} & \numprint{39,0}   & \numprint{41,8}    & \numprint{47,4}    & \numprint{28,6}   & \numprint{31,6}    & \numprint{37,0}\\
\texttt{\detokenize{bmwcra_1}}      & \numprint{0,3}  & \numprint{4,3} & \numprint{149,8}  & \numprint{206,0}   & \numprint{216,5}   & \numprint{62,9}   & \numprint{110,1}   & \numprint{151,4}\\
\texttt{\detokenize{boneS01}}        & \numprint{0,3}  & \numprint{7,1} & \numprint{222,1}  & \numprint{245,7}   & \numprint{258,1}   & \numprint{55,7}   & \numprint{75,0}    & \numprint{96,8}\\
\texttt{\detokenize{brack2}}         & \numprint{0,1}  & \numprint{0,1} & \numprint{10,1}   & \numprint{15,1}    & \numprint{20,2}    & \numprint{5,5}    & \numprint{11,1}    & \numprint{15,4}\\
\texttt{\detokenize{cfd2}}           & \numprint{0,2}  & \numprint{0,2} & \numprint{73,3}   & \numprint{103,8}   & \numprint{114,2}   & \numprint{27,5}   & \numprint{62,4}    & \numprint{77,5}\\
\texttt{\detokenize{cont-300}}       & \numprint{0,1}  & \numprint{0,1} & \numprint{12,8}   & \numprint{25,9}    & \numprint{41,3}    & \numprint{7,8}    & \numprint{23,0}    & \numprint{33,0}\\
\texttt{\detokenize{cop20k_A}}      & \numprint{0,2}  & \numprint{1,5} & \numprint{69,1}   & \numprint{88,0}    & \numprint{100,9}   & \numprint{18,0}   & \numprint{39,6}    & \numprint{51,1}\\
\texttt{\detokenize{crack}}          & \numprint{0,1}  & \numprint{0,1} & \numprint{2,0}    & \numprint{3,1}     & \numprint{3,6}     & \numprint{1,6}    & \numprint{2,8}     & \numprint{3,2}\\
\texttt{\detokenize{cs4}}            & \numprint{0,1}  & \numprint{0,1} & \numprint{5,6}    & \numprint{8,4}     & \numprint{9,4}     & \numprint{4,3}    & \numprint{7,5}     & \numprint{8,3}\\
\texttt{\detokenize{cti}}            & \numprint{0,1}  & \numprint{0,1} & \numprint{5,3}    & \numprint{5,9}     & \numprint{6,7}     & \numprint{3,7}    & \numprint{4,4}     & \numprint{5,3}\\
\texttt{\detokenize{data}}           & \numprint{0,1}  & \numprint{0,1} & \numprint{1,8}    & \numprint{2,1}     & \numprint{2,4}     & \numprint{1,6}    & \numprint{1,9}     & \numprint{2,2}\\
\texttt{\detokenize{del23}}          & \numprint{7,9}  & \numprint{3,6} & \numprint{1154,2} & \numprint{4114,6}  & \numprint{6362,4}  & \numprint{1306,3} & \numprint{4077,3}  & \numprint{6159,8}\\
\texttt{\detokenize{del24}}          & \numprint{17,4} & \numprint{7,2} & \numprint{2733,4} & \numprint{12807,8} & \numprint{18613,6} & \numprint{2580,3} & \numprint{12711,2} & \numprint{17219,3}\\
\texttt{\detokenize{deu}}            & \numprint{4,8}  & \numprint{1,3} & \numprint{337,6}  & \numprint{860,1}   & \numprint{1032,3}  & \numprint{275,7}  & \numprint{906,4}   & \numprint{1086,2}\\
\texttt{\detokenize{Dubcova3}}       & \numprint{0,2}  & \numprint{1,0} & \numprint{42,2}   & \numprint{65,4}    & \numprint{82,2}    & \numprint{18,3}   & \numprint{42,2}    & \numprint{59,7}\\
\texttt{\detokenize{eur}}            & \numprint{24,0} & \numprint{5,3} & \numprint{2117,7} & \numprint{8213,9}  & \numprint{8748,8}  & \numprint{2135,6} & \numprint{8921,6}  & \numprint{9323,0}\\
\texttt{\detokenize{fe_4elt2}}      & \numprint{0,1}  & \numprint{0,1} & \numprint{1,4}    & \numprint{1,6}     & \numprint{1,9}     & \numprint{1,2}    & \numprint{1,5}     & \numprint{1,8}\\
\texttt{\detokenize{fe_body}}       & \numprint{0,1}  & \numprint{0,1} & \numprint{5,8}    & \numprint{8,9}     & \numprint{8,6}     & \numprint{4,6}    & \numprint{8,0}     & \numprint{7,8}\\
\texttt{\detokenize{fe_ocean}}      & \numprint{0,1}  & \numprint{0,2} & \numprint{12,3}   & \numprint{23,4}    & \numprint{34,9}    & \numprint{7,9}    & \numprint{20,1}    & \numprint{33,4}\\
\texttt{\detokenize{fe_pwt}}        & \numprint{0,1}  & \numprint{0,1} & \numprint{4,0}    & \numprint{6,1}     & \numprint{7,4}     & \numprint{2,8}    & \numprint{5,3}     & \numprint{7,3}\\
\texttt{\detokenize{fe_rotor}}      & \numprint{0,1}  & \numprint{0,3} & \numprint{27,2}   & \numprint{39,2}    & \numprint{47,3}    & \numprint{9,9}    & \numprint{22,6}    & \numprint{33,7}\\
\texttt{\detokenize{fe_sphere}}     & \numprint{0,1}  & \numprint{0,1} & \numprint{3,1}    & \numprint{4,0}     & \numprint{4,6}     & \numprint{2,0}    & \numprint{3,3}     & \numprint{4,2}\\
\texttt{\detokenize{fe_tooth}}      & \numprint{0,1}  & \numprint{0,2} & \numprint{41,0}   & \numprint{68,9}    & \numprint{74,7}    & \numprint{14,4}   & \numprint{45,7}    & \numprint{60,1}\\
\texttt{\detokenize{finan512}}       & \numprint{0,1}  & \numprint{0,1} & \numprint{7,1}    & \numprint{9,8}     & \numprint{12,8}    & \numprint{5,2}    & \numprint{8,5}     & \numprint{12,1}\\
\texttt{\detokenize{G2_circuit}}    & \numprint{0,1}  & \numprint{0,1} & \numprint{13,5}   & \numprint{19,7}    & \numprint{24,6}    & \numprint{8,4}    & \numprint{16,9}    & \numprint{24,8}\\
\texttt{\detokenize{m14b}}           & \numprint{0,3}  & \numprint{0,3} & \numprint{60,1}   & \numprint{90,3}    & \numprint{114,3}   & \numprint{28,6}   & \numprint{60,6}    & \numprint{78,7}\\
\texttt{\detokenize{memplus}}        & \numprint{0,1}  & \numprint{0,3} & \numprint{32,5}   & \numprint{37,2}    & \numprint{32,8}    & \numprint{24,6}   & \numprint{30,8}    & \numprint{27,4}\\
\texttt{\detokenize{nlr}}            & \numprint{7,5}  & \numprint{4,0} & \numprint{407,1}  & \numprint{1935,6}  & \numprint{3217,7}  & \numprint{320,5}  & \numprint{1940,9}  & \numprint{3085,0}\\
\texttt{\detokenize{rgg23}}          & \numprint{9,7}  & \numprint{4,2} & \numprint{2088,1} & \numprint{6434,3}  & \numprint{7651,5}  & \numprint{2239,4} & \numprint{7493,0}  & \numprint{8127,6}\\
\texttt{\detokenize{rgg24}}          & \numprint{21,5} & \numprint{9,0} & \numprint{3116,0} & \numprint{9616,8}  & \numprint{10415,3} & \numprint{2963,8} & \numprint{9530,6}  & \numprint{10436,2}\\
\texttt{\detokenize{shipsec5}}       & \numprint{0,3}  & \numprint{3,0} & \numprint{114,2}  & \numprint{146,9}   & \numprint{177,2}   & \numprint{46,9}   & \numprint{90,8}    & \numprint{129,0}\\
\texttt{\detokenize{t60k}}           & \numprint{0,1}  & \numprint{0,1} & \numprint{2,2}    & \numprint{6,4}     & \numprint{8,6}     & \numprint{1,7}    & \numprint{6,5}     & \numprint{8,1}\\
\texttt{\detokenize{thermal2}}       & \numprint{0,9}  & \numprint{0,5} & \numprint{68,3}   & \numprint{320,5}   & \numprint{638,1}   & \numprint{61,8}   & \numprint{326,0}   & \numprint{622,4}\\
\texttt{\detokenize{thermomech}}     & \numprint{0,1}  & \numprint{0,1} & \numprint{5,0}    & \numprint{22,9}    & \numprint{27,9}    & \numprint{3,6}    & \numprint{20,1}    & \numprint{26,1}\\
\texttt{\detokenize{uk}}             & \numprint{0,1}  & \numprint{0,1} & \numprint{0,9}    & \numprint{1,1}     & \numprint{1,3}     & \numprint{0,8}    & \numprint{1,0}     & \numprint{1,2}\\
\texttt{\detokenize{vibrobox}}       & \numprint{0,1}  & \numprint{0,8} & \numprint{44,7}   & \numprint{47,9}    & \numprint{48,0}    & \numprint{21,5}   & \numprint{25,3}    & \numprint{25,3}\\
\texttt{\detokenize{wave}}           & \numprint{0,2}  & \numprint{0,3} & \numprint{118,2}  & \numprint{157,5}   & \numprint{183,7}   & \numprint{28,2}   & \numprint{71,5}    & \numprint{94,6}\\
\texttt{\detokenize{whitaker3}}      & \numprint{0,1}  & \numprint{0,1} & \numprint{1,4}    & \numprint{2,2}     & \numprint{2,5}     & \numprint{1,2}    & \numprint{2,0}     & \numprint{2,3}\\
\texttt{\detokenize{wing}}           & \numprint{0,1}  & \numprint{0,1} & \numprint{14,5}   & \numprint{25,7}    & \numprint{29,6}    & \numprint{8,2}    & \numprint{19,8}    & \numprint{23,9}\\
\texttt{\detokenize{wing_nodal}}    & \numprint{0,1}  & \numprint{0,1} & \numprint{9,0}    & \numprint{9,2}     & \numprint{10,7}    & \numprint{5,8}    & \numprint{6,4}     & \numprint{8,2}\\
\bottomrule
\end{tabular}
\vspace*{.5cm}
\caption{Detailed per instances results as average running time.}                
\label{tab:detailedtime}
\end{table}
\vfill
\pagebreak
\end{appendix}                                         
\end{document}